\documentclass[usenatbib]{mn2e}
\usepackage{psfig}

\title[Roche tomography of CVs -- II. Images of the secondary stars in CVs]
{Roche tomography of cataclysmic variables -- II. Images of the
secondary stars in AM Her, QQ Vul, IP Peg and HU Aqr}

\author[C.\,A.\ Watson, V.\,S.\ Dhillon, R.\,G.\,M.\ Rutten, and
A.\,D.\ Schwope] {C.\,A.\ Watson,$^1$\thanks{E-mail:
c.watson@sheffield.ac.uk} V.\,S.\ Dhillon,$^1$ R.\,G.\,M.\ Rutten,$^2$
A.\,D.\ Schwope$^3$\\ $^1$Department of Physics and Astronomy,
University of Sheffield,  Sheffield S3 7RH, UK \\ $^2$Isaac Newton
Group, Apartado de Correos 321, 38780 Santa Cruz de La Palma, Canary
Islands, Spain\\ $^3$Astrophysikalisches Institut Potsdam, An der
Sternwarte 16, 14482 Potsdam, Germany\\}

\date{\center{\Large Accepted for publication in the Monthly 
Notices of the Royal Astronomical Society \\ 
\vspace{.5cm} 6th February 2003}} 

\begin{document}
\maketitle

\begin{abstract}
We present a set of Roche tomography reconstructions of the secondary
stars in the cataclysmic variables AM Her, QQ Vul, IP Peg and HU Aqr.
The image reconstructions show distinct asymmetries in the irradiation
pattern for all four systems which can be attributed to shielding of
the secondary star by the accretion stream/column in AM Her, QQ Vul
and HU Aqr, and increased irradiation by the bright spot in IP Peg.
We use the entropy landscape technique to derive accurate system
parameters ($M{_1}$, $M_2$, $i$ and $\gamma$) for the four binaries.
In principle, this technique should provide the most reliable mass
determinations available, since the intensity distribution across the
secondary star is known. We also find that the intensity distribution
can systematically affect the value of $\gamma$ derived from circular
orbit fits to radial velocity variations.
\end{abstract} 


\section{Introduction}
\label{sec:introduction}

The secondary, Roche-lobe-filling stars in CVs are key to our
understanding of the origin, evolution and behaviour of this class of
interacting binary. To best study the secondary stars in CVs, we would
ideally like direct images of the stellar surface. This is currently
impossible, however, as typical CV secondary stars have radii of  400
000 km and distances of 200 pc, which means that to detect a feature
covering 20 per cent of the star's surface requires a resolution of
approximately 1 microarcsecond, 10 000 times greater than the
diffraction-limited resolution of the world's largest
telescopes. \citet{rutten94} and \citet[hereafter referred to as Paper
I]{watson01} described a way around this problem using an indirect
imaging technique called {\em Roche tomography} which uses
phase-resolved spectra to reconstruct the line intensity distribution
on the surface of the secondary star.

Obtaining surface images of the secondary star in CVs has far-reaching
implications.  For example, a knowledge of the irradiation pattern on
the inner hemisphere of the secondary star in CVs is essential if one
is to calculate stellar masses accurately enough to test binary star
evolution models (see \citealt{smith98}). Furthermore, the irradiation
pattern provides information on the geometry of the accreting
structures around the white dwarf (see \citealt{smith95}).

In this paper we present new Roche tomograms of the secondary star in
the dwarf nova IP Peg, and the magnetic CVs AM Her and QQ Vul in the
light of the Na I $\lambda\lambda$8183,8195 \AA~absorption doublet. In
addition, we also present a Roche tomogram of the magnetic CV HU Aqr
in the light of the He II $\lambda$4686 \AA~emission line component
known to originate from the secondary star.  These tomograms allow a
study of the irradiation pattern on the secondary stars as well as
providing measurements of the binary parameters for all four CVs.

\section{Observations and reduction}
\label{sec:observations}

The spectra of QQ Vul and HU Aqr were taken on the 3.5-m telescope at
Calar Alto and full details of the observations and subsequent data
reduction can be found in \citet*{catalan99} and \citet*{schwope97},
respectively.

The observations of IP Peg and AM Her were carried out over two nights
on 1994 August 11--12 using the ISIS dual-beam spectrograph and 1200
lines mm$^{-1}$ gratings on the 4.2m William Herschel Telescope
(WHT). The EEV3 CCD chip in the red arm covered the wavelength range
7985--8437\AA~at a resolution of 0.37\AA~ ($\sim$27 km s$^{-1}$).  The
TEK2 CCD chip on the blue arm covered the wavelength range 4560--4968\AA~
at a resolution of 0.4\AA~($\sim$50 km s$^{-1}$).  To reduce readout
noise, the pixels were binned by a factor of 2 in the spatial
direction. Each object was observed using exposure times of 300
seconds and a nearby field star was located on the slit to correct for
slit losses. Wide slit spectra of the companion star were also taken
to permit absolute spectrophotometry. Sky-flats and tungsten flats
were taken on both nights and a CuAr+CuNe arc spectrum was taken
approximately every 50 minutes in order to wavelength calibrate the
spectra.  Narrow and wide slit spectra at a range of different
air-masses were taken of flux standards for the purposes of flux
calibration and water removal. We also observed a range of M-dwarf
spectral-type templates in order to obtain a Na I intrinsic line
profile for use in Roche tomography.

During the reduction process, bias subtraction, division by a master
flat-field frame and sky subtraction were carried out and then the
spectra were optimally extracted \citep{horne86a}. The arc spectra were
used to wavelength calibrate the data, with an rms error of
0.01\AA. Slit loss correction was then applied by dividing the object
spectra by the average value per pixel (after masking out strong
lines, cosmic rays and telluric features) in the corresponding
comparison star spectra, and later multiplying by the average value
per pixel in the wide-slit comparison star spectrum. Telluric
absorption and absolute flux calibration were then performed using the
standard star spectra.

In total we obtained 79 spectra of IP Peg and 51 spectra of AM Her. We
present only the red spectra in this paper, as we are only concerned
with the Na I doublet at $\lambda\lambda$8183,8195\AA~ for the
purposes of this work.  A full journal of observations is presented in
Table~\ref{tab:journal}.

\begin{table*}
\caption[]{Journal of observations. Orbital phase is shown for AM Her
using the ephemeris of \protect\citet{southwell95}, and IP Peg using
the ephemeris of \protect\citet{wolf93}. Since the latter ephemeris is
deduced from the white-dwarf egress in IP Peg, an appropriate
correction has been made to ensure that phase 0 corresponds to the
superior conjunction of the primary.  Note: HU Aqr was trailed along
the slit during the exposures.  The final time resolution was $\sim$
30 s, and the data was phase-folded into 100 phase bins covering a
complete orbital cycle.}  \centering
\begin{tabular}{llccccccc}
\hline
Object & UT Date & UT & UT & Phase & Phase & $T_{\exp}$ & No. of \\
& (D/M/Y) & start & end & start & end & (s) & Spectra \\ \hline
AM Her & 11/08/94 & 21:38 & 23:58 & 0.57 & 1.29 & 300 & 22 \\
AM Her & 12-13/08/94 & 21:20 & 00:09 & 2.22 & 3.11 & 300 & 29 \\
QQ Vul & 08-09/07/91 & 23:33 & 04:10 & 0.12 & 1.33 & 345--720 & 16 \\
QQ Vul & 09-10/07/91 & 21:47 & 03:52 & 6.12 & 7.71 & 600--680 & 20 \\
IP Peg & 12/08/94 & 01:23 & 05:26 & 0.88 & 1.93 & 300 & 41 \\
IP Peg & 13/08/94 & 01:13 & 05:11 & 7.16 & 8.18 & 300 & 38 \\
HU Aqr & 17-18/08/93 & & & & & 750--1800 & 7 \\
\hline \\
\label{tab:journal}
\end{tabular}
\end{table*}

\section{Roche tomography}

\subsection{Techniques}
Roche tomography (\citealt{rutten94}; \citealt{dhillon01};
Paper I) is analogous to the Doppler imaging technique used
to map rapidly rotating single stars (e.g. \citealt{vogt83}). In Roche
tomography the secondary star is assumed to be Roche-lobe-filling,
locked in synchronous rotation and to have a circularized orbit.  The
secondary star is then modelled as a series of quadrilateral tiles or
surface elements, each of which is assigned a copy of the local
(intrinsic) specific intensity profile. These profiles are then scaled
to take into account the projected area of the surface element, limb
darkening and obscuration, and then Doppler shifted according to the
radial velocity of the surface element at that particular
phase. Summing up the contributions from each element gives the
rotationally broadened profile at that particular orbital phase.

By iteratively varying the strengths of the profile contributed from
each element, the `inverse' of the above procedure can be performed. Due
to the variable and unknown contribution to the spectrum of the
accretion regions in CVs, the data must be slit-loss corrected and
continuum subtracted prior to mapping with Roche tomography. Thus,
Roche tomograms present images of the distribution of line flux on the
secondary star. The data is fit to an aim $\chi^2$, and a unique
map is selected by employing the maximum entropy (MEMSYS) algorithm
developed by \citet{skilling84}. The definition of image entropy, $S$,
that we use is
\[S = \sum_{j=1}^k m_j - d_j - m_j \ln \left(m_j/d_j \right) \]
where $k$ is the number of tiles in the map, and $m_j$ and $d_j$ are
the map of line flux and the default map, respectively. We employ a
moving uniform default map, where each element in the default is set
to the average value in the map. For further details of the Roche
tomography process see the review by \citet{dhillon01}.

For those CVs where we wish to map the Na I doublet, Roche tomography
requires the intrinsic line profile to be input as a
template. Ideally, this should be the same line taken from the
spectrum of a slowly rotating star of the same spectral class as the
star being mapped, preferably obtained with the same instrumental
set-up. Unfortunately, only one of the spectral-type template stars
that were observed is useful (GL806 -- see Table~\ref{tab:templates});
the others show signs of broadening of the Na I doublet and have been
identified as either flare stars, and are therefore probably rapidly
rotating, or binary stars, where $V_{rot}\sin i$ measurements using
the Na I absorption doublet have been found to be systematically
greater than those given by the narrow metal lines
(\citealt{bleach00}). We used the GL806 template in all of our
reconstructions and, even though the spectral class of the template
differs from those of the stars being mapped, the exact shape of the
template profile used has little effect on the reconstructions due to
the large degree of rotational broadening present in CV secondary
stars.

For the CVs mapped using the Na I doublet, it is necessary to correct
the derived systemic velocities for the systemic velocity of the
spectral-type template used in the reconstructions. The systemic
velocity of GL806 was determined by fitting gaussians to the Na I
doublet, which yielded a value of $\gamma$ = --24.3 $\pm$ 1.1 km
s$^{-1}$. For comparison, \citet{wilson53} found the systemic velocity
of GL806 to be --15 $\pm$ 5 km s$^{-1}$, whereas \citet{gliese69}
quote a value of --24.4 km s$^{-1}$.

It is important to note that the maximum-entropy algorithm we use
requires that the input data are positive. As a result, it is
necessary to invert absorption-line profiles and, in the case of low
signal-to-noise data, add a positive constant in order to ensure that
there are no negative values in the continuum-subtracted data. If this
positive constant is not added, and any negative points are either set
to zero or ignored during the reconstruction, then the fit will be
positively biased in the wings of the line profile wings. The exact
value of this constant is unimportant, so long as it prevents the
occurrence of negative values in the data and is effectively
removed. We account for this positive constant during the
reconstruction by employing a virtual image element which contributes
a single value to all data points, effectively cancelling out the
constant.  For the systems where the Na I doublet has been used in the
reconstructions, we have employed this technique of adding a positive
constant to each data value. We find that this significantly improves
the reliability of the technique to determine the component masses
when noisy data are being used. As the data for HU Aqr are from
Gaussian fits to the He II $\lambda$4686\AA~ emission line, there are
no errors in the continuum and hence we do not apply this technique.

In addition, we have accounted for the effects of velocity smearing due
to the motion of the secondary star over the duration of an exposure for
all four systems. This is done by calculating the profile at
evenly spaced intervals over one exposure and then averaging them.
Although computationally more time-consuming, this reduces the effects
of smearing, and can also take into account features that
may cross the limb of the secondary star during an exposure.

\subsection{Determining the system parameters}
\label{sec:parameters}

Roche tomograms are greatly influenced by parameters such as
the component masses, inclination and systemic velocity of the binary.
Errors in the assumed values of these parameters degrades the quality
of the surface map and, in general, results in additional structure in
the Roche tomograms. The correct parameters are, therefore, those that
produce the map containing the least artefacts, corresponding to the
map of highest entropy (Paper I). By carrying out
reconstructions for many pairs of component masses (iterating to the
same $\chi^2$ on each occasion) we can construct an `entropy
landscape' (e.g., Fig.~\ref{fig:landscapes}), where each point
corresponds to the entropy value obtained in a reconstruction
for a particular pair of component masses.

\begin{figure*}
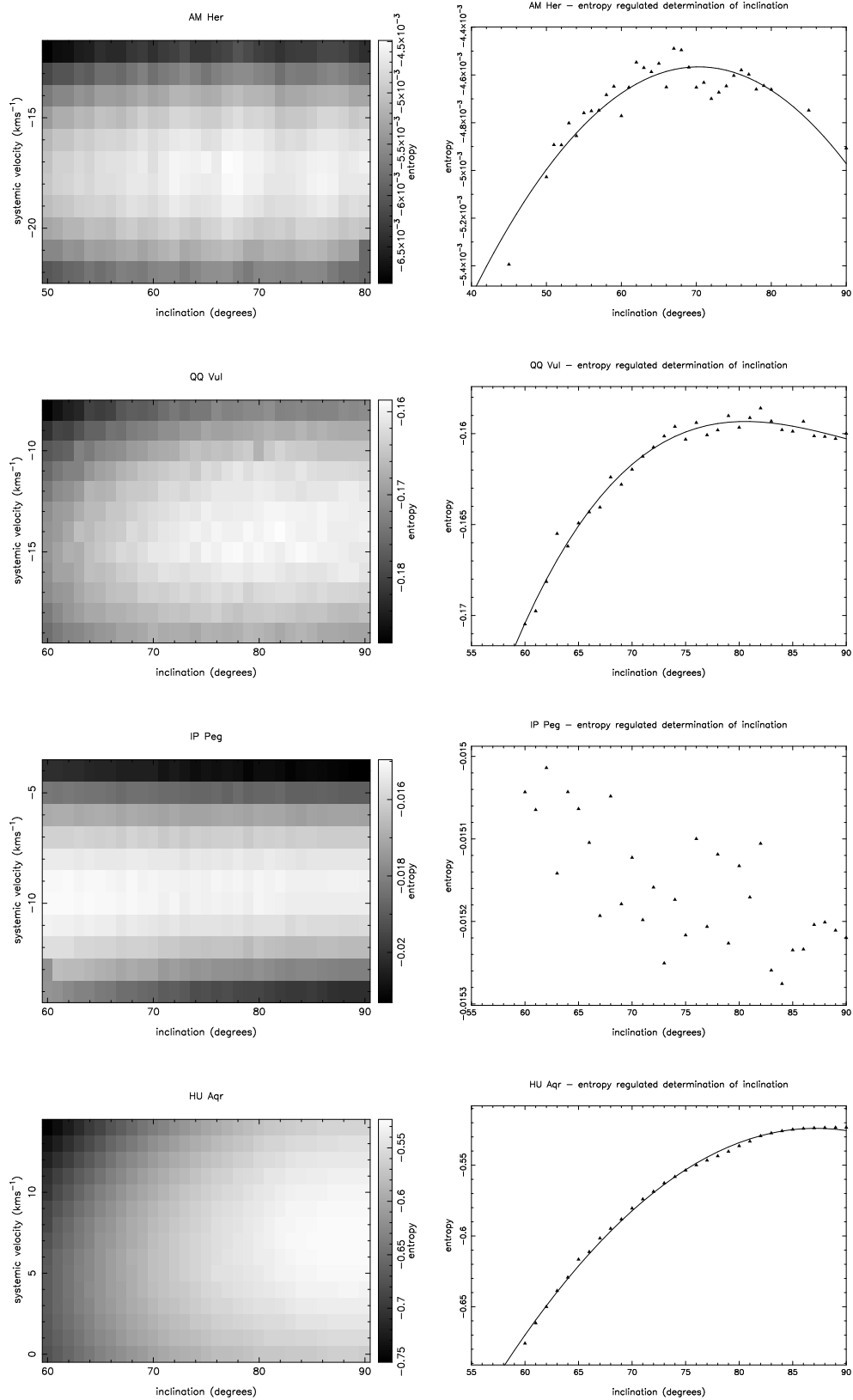

\begin{tabular}{ll}
\psfig{figure=fig1a.ps,width=6.3cm,angle=-90} &
\psfig{figure=fig1b.ps,width=6.3cm,angle=-90} \\
\vspace{0.2cm}\\
\psfig{figure=fig1c.ps,width=6.3cm,angle=-90} &
\psfig{figure=fig1d.ps,width=6.3cm,angle=-90} \\
\vspace{0.2cm}\\
\psfig{figure=fig1e.ps,width=6.3cm,angle=-90} &
\psfig{figure=fig1f.ps,width=6.3cm,angle=-90} \\
\vspace{0.2cm}\\
\psfig{figure=fig1g.ps,width=6.3cm,angle=-90} &
\psfig{figure=fig1h.ps,width=6.3cm,angle=-90} \\
\vspace{0.2cm}\\\end{tabular}
\caption{The plots on the left show the maximum entropy value obtained
in each entropy landscape when a particular combination of inclination
and systemic velocity is employed during the reconstructions;
horizontal cuts through these plots at the optimal systemic velocity
found for each object are shown on the right. From top to bottom: AM
Her, QQ Vul, IP Peg and HU Aqr. The solid curves show parabolic fits
to the data points.}
\label{fig:incl}
\end{figure*}

In order to determine the binary parameters of the CVs studied in this
paper, we constructed a series of entropy landscapes assuming a wide
range of different values for the orbital inclination and systemic
velocity. For each combination of $i$ and $\gamma$, we picked the
maximum entropy value in the corresponding entropy landscape, and
plotted this value in the $i-\gamma$ plane, which is shown for each
object on the left-hand side of Fig.~\ref{fig:incl}. In all cases we
found that a unique or optimal systemic velocity could be selected
that consistently gave the map of maximum entropy. In addition, the
value of the optimal systemic velocity was found to be largely
independent of the orbital inclination assumed during the
reconstructions, as demonstrated in the left-hand plots of
Fig.~\ref{fig:incl}. For example, in reconstructions carried out over a
range of inclinations spanning 30$^{\circ}$, the systemic velocity we
obtain for both IP Peg and AM Her only varies by 1 km s$^{-1}$.

The inclination, however, is not as well constrained for any of
the CVs. The right-hand plots on Fig.~\ref{fig:incl} show cuts through
the plots in the $i-\gamma$ plane at the optimal systemic velocity,
and the scatter clearly demonstrates the difficulty in selecting an
optimal inclination, although the deterioration in the quality of
the reconstruction for some inclinations allow limits to be set.

\begin{center}
\begin{table}
\caption[]{Spectral-type templates observed with the WHT.}
\begin{tabular}{lll}
\hline
Object & Spec. Type & Notes\\ \hline
GL 65B & M5.5V & flare star \\
GL 866 & M5.5V & flare star \\
GL 83.1A & M4.5V & flare star \\
GL 699 & M4Ve & variable star\\
GL 725A & M3V & double star \\
GL 806 & M1.5V & used in this paper \\
\hline \\
\label{tab:templates}
\end{tabular}
\end{table}
\end{center}

The difficulty in constraining the inclination can be explained by
considering that there are two ways of determining the inclination of
non-eclipsing CVs from the spectra of the secondary star alone. First,
from the variation in the projected radius of the Roche-lobe shaped
secondary star as it is viewed at different aspects, corresponding to
a variation in the shape (or measured $v\sin i$) of the line profiles
(see \citealt{shahbaz98}). The second is from the variation in the
strength of the profiles, since a high inclination system will exhibit
a much greater variation in the line profile strength than a low
inclination system with the same surface intensity distribution.

Unfortunately, the variation in $v\sin i$ is expected to be, at most,
around 20--25 km s$^{-1}$ for high inclination systems. Although this
is comparable to the velocity resolution of our data, it is buried in
noise and hence the mapping technique is largely blind to any such
variations.  Our only constraint on the inclination, therefore, comes
from the variation of the line strength, which also has its
limitations. Although, rapidly varying features can only be mapped
onto high inclination systems, slowly varying features produced by a
low inclination system can also be mapped onto the polar regions of a
high inclination system with little loss in the quality of the
reconstruction. Thus we only expect to constrain the inclination in
CVs which have high inclinations and exhibit strong asymmetries in
their surface intensity distributions. In addition, for the same reason
outlined above, we also expect our determinations to be biased towards
higher inclinations.

\begin{figure*}
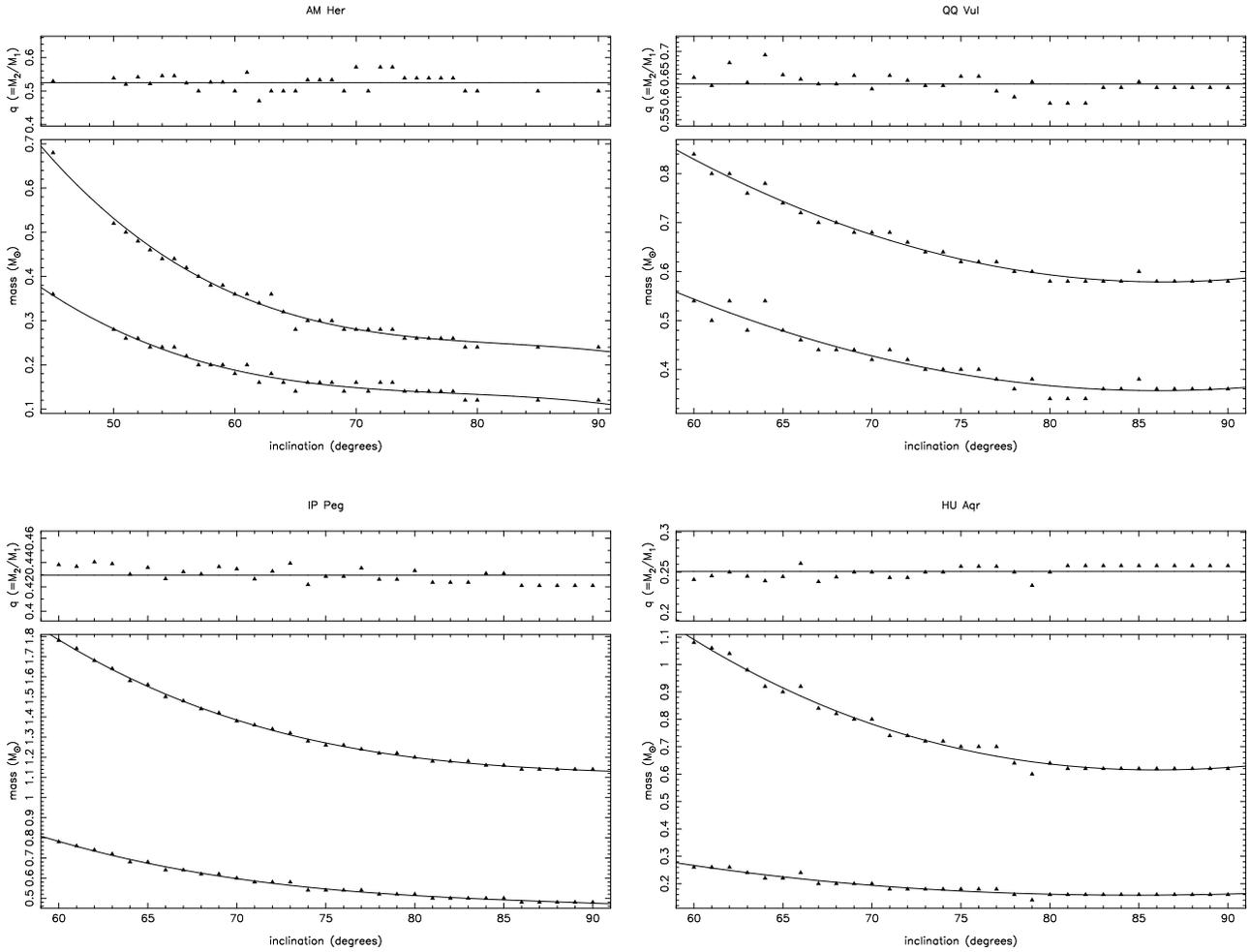

\begin{tabular}{ll}
\psfig{figure=fig2a.ps,width=8.3cm,angle=-90} &
\psfig{figure=fig2b.ps,width=8.3cm,angle=-90} \\
\vspace{0.2cm}\\
\psfig{figure=fig2c.ps,width=8.3cm,angle=-90} &
\psfig{figure=fig2d.ps,width=8.3cm,angle=-90} \\
\vspace{0.2cm}\\\end{tabular}
\caption{Estimates of the component masses for different inclinations.
Clockwise from top left: AM Her, QQ Vul, HU Aqr and IP Peg. 
The top panel of each plot shows the mass ratio, the lower panel shows the
individual masses with the upper curve showing the primary mass and the lower
curve showing the secondary mass.}
\label{fig:mratio}
\end{figure*}

\begin{figure*}
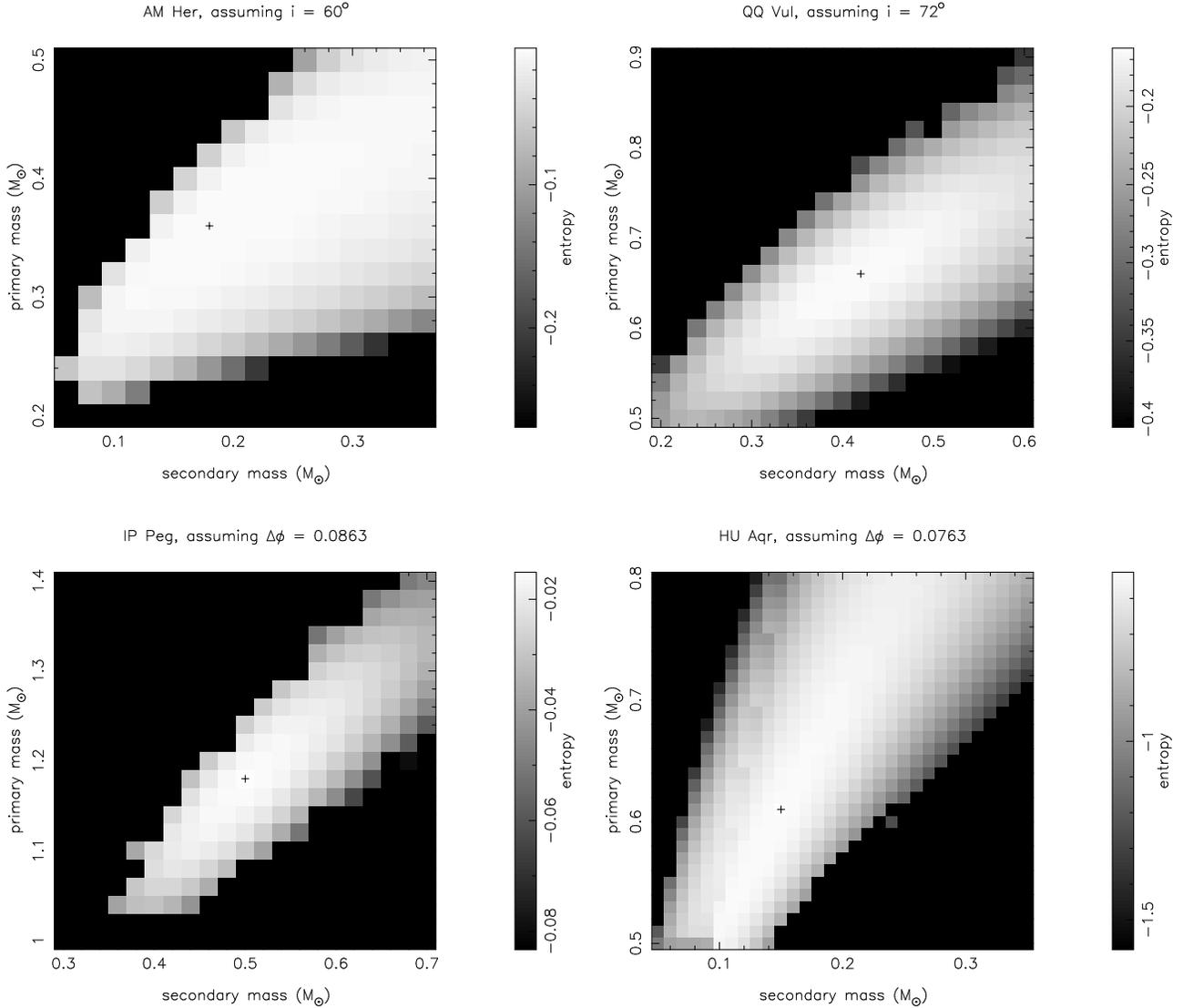

\begin{tabular}{ll}
\psfig{figure=fig3a.ps,width=8.3cm,angle=-90} &
\psfig{figure=fig3b.ps,width=8.3cm,angle=-90} \\
\vspace{0.2cm}\\
\psfig{figure=fig3c.ps,width=8.3cm,angle=-90} &
\psfig{figure=fig3d.ps,width=8.3cm,angle=-90} \\
\vspace{0.2cm}\\\end{tabular}
\caption{Entropy landscapes for the four CVs. Clockwise from top left:
AM Her, QQ Vul, HU Aqr and IP Peg. Dark regions indicate masses for which no
acceptable fit could be found. The cross marks the point of maximum entropy
in each landscape.}
\label{fig:landscapes}
\end{figure*}

As we cannot reliably constrain the inclination of the non-eclipsing
CVs in this work, and hence the component masses,
Fig.~\ref{fig:mratio} shows the derived component masses as a function
of inclination. The top section of each panel in Fig.~\ref{fig:mratio}
also shows the mass ratio derived for each inclination. Since $v\sin
i$ is purely a function of the mass ratio $q$ (=$M_1/M_2$) and the
radial velocity semi-amplitude of the secondary star $K_r$,
reconstructions over a range of inclinations should yield the same
value of the mass ratio on each occasion. This is confirmed in
Fig.~\ref{fig:mratio}, with the scatter in the values serving as an
indication of the noise in the technique. Indeed, the scatter can be
accounted for by the fact that the component masses are only varied in
0.02M$_{\odot}$ increments in the entropy landscapes.

Entropy landscapes representative of the  mass estimates for each
system are shown in Fig.~\ref{fig:landscapes}, and a summary of the
adopted system parameters can be found in
Table~\ref{tab:estimates}. The results for each CV are discussed in
more detail in the relevant sections below.

\begin{center}
\begin{table}
\caption[]{Adopted system parameters.}
\begin{tabular}{lcccc} \hline
& AM Her & QQ Vul & IP Peg & HU Aqr \\ \hline
$i$ & 60--80$^{\circ}$ & 72$^{\circ}\leq$~i~$\leq$90$^{\circ}$ & 82$^{\circ}$--
85$^{\circ}$ & 84.4$^{\circ}$ \\
$q$ & 0.53 & 0.63 & 0.43 & 0.25\\
$M_1$ ($M_{\odot}$) & $\sim$0.36 & 0.58--0.66 & 1.16--1.18 & 0.61\\
$M_2$ ($M_{\odot}$)& $\sim$0.18 & 0.34--0.44& 0.50 & 0.15\\
$\gamma$ (km s$^{-1}$) &--17 & --14 & --9 & +7\\
\hline
\label{tab:estimates}
\end{tabular}
\end{table}
\end{center}

\section{Results for AM Her}
\label{sec:amher}

\subsection{The binary geometry of AM Her}

From Fig.~\ref{fig:incl} the maximum entropy value yields a value of
--17 km s$^{-1}$ for the systemic velocity of AM Her, taking into
account the systemic velocity of the template star. This is within
2$\sigma$ of the range of values found by \citet*{young81b} of
--14 $\pm$ 4 km s$^{-1}$ and --40 $\pm$ 15 km s$^{-1}$ from circular fits
to the H$\gamma$ and He I $\lambda$4471\AA~emission lines,
respectively.

Fig.~\ref{fig:incl} also shows the maximum entropy value as a function
of inclination obtained in entropy landscapes constructed assuming an
optimum systemic velocity of --17 km s$^{-1}$. It appears as though we
cannot determine a unique inclination for AM Her, though our work
suggests that the inclination may lie around 60--80$^{\circ}$. Given
that we expect our results to be biased towards higher inclinations,
the inclination is most likely to be around 60$^{\circ}$.  Although
this estimate is far greater than the $\sim$35$^{\circ}$ determined by
\citet{Brainerd85}, it is consistent with the findings of
\citet{davey96} who used radial-velocity curve fitting to constrain
the inclination to be between 45$^{\circ}$ and 60$^{\circ}$.
\citet{greeley99} observed a partial eclipse of the He II
$\lambda$1640 \AA~emission line region by the secondary star and hence
inferred an inclination $\ge$45$^{\circ}$. In addition to this,
\citet{Wickramasinghe91c} obtained $i = 52^{\circ}$ from polarimetric
observations.

Fig.~\ref{fig:landscapes} shows the entropy landscape for AM Her
constructed using $i$ = 60$^{\circ}$, $\gamma$ = --17 km s$^{-1}$ and
taking into account velocity smearing introduced by the 300 second
exposure times.  The brightest point on this plot indicates the map of
highest entropy, and corresponds to $M_1$ = 0.36 M$_{\odot}$
and $M_2$ = 0.18 M$_{\odot}$, which we adopt as our best
estimate of the component masses. The optimum component masses
decrease to $M_1$ = 0.24 M$_{\odot}$ and $M_2$ = 0.12 M$_{\odot}$ for
$i$~=~80$^{\circ}$.  These masses appear suspiciously low since AM Her
lies above the period gap and we would not, therefore, expect the
secondary mass to be significantly less than
$M_2$~=~$\sim$0.25~M$_{\odot}$.  This partly confirms our speculation
that  our inclination determinations are biased towards high
inclinations (Section~\ref{sec:parameters}) and that the inclination
is not likely to be greater than 60$^{\circ}$.

The mass of the secondary star derived from the entropy landscape
agrees well with the value of $M_2$ = 0.20 -- 0.26 M$_{\odot}$
estimated by \citet{southwell95}.  This also agrees with the
main-sequence mass-period relation derived by \citet{smith98}, which
gives a secondary star mass of 0.23$\pm$0.02 M$_{\odot}$.

Meaningful comparison of the primary mass derived in our work with
other published white dwarf masses is, unfortunately, more difficult
as they cover a wide range of values, including: 0.39 M$_{\odot}$
\citep{young81b}, 0.69 M$_{\odot}$ \citep*{wu95}, 0.75 M$_{\odot}$
\citep{mukai87}, 0.91 M$_{\odot}$ \citep{mouchet93} and 1.22 M$_{\odot}$
\citep*{cropper98}. \citet{gansicke98}, however, concluded that the
white dwarf mass in AM Her was between 0.35 -- 0.53 M$_{\odot}$ and
ruled out a white dwarf mass as high as 1.22 M$_{\odot}$. In addition
to their secondary star mass estimate, \citet{southwell95} derived $q$
= 0.47$\pm$0.08, in agreement with the value of $q$ = 0.53
obtained in this work (Fig.~\ref{fig:mratio}).

\subsection{The surface map of AM Her}
\label{amhermap}

\begin{figure*}
\psfig{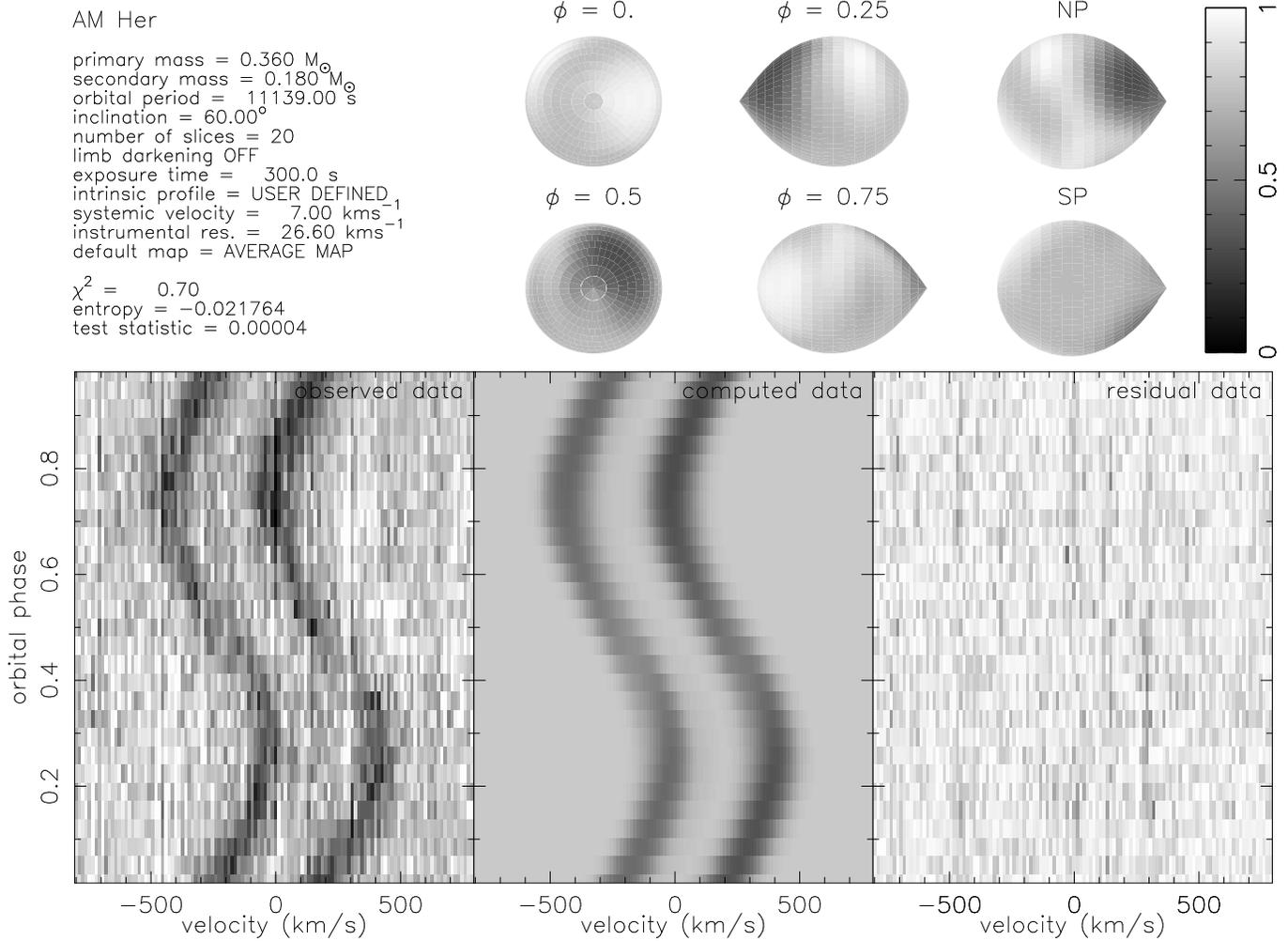}
\caption{Roche tomogram of AM Her. Bright greyscales in the tomograms
depict regions of enhanced Na I absorption. The greyscales are
reversed in the trailed spectra.  Note that the systemic velocity
quoted in the upper-left of the panel does not take into account the
systemic velocity of the spectral-type template star.}
\label{fig:amherdisp}
\end{figure*}

The Roche tomogram for AM Her is presented in Fig.~\ref{fig:amherdisp}
and shows a decrease in the amount of Na I absorption around the inner
Lagrangian ($L_1$) point on the trailing hemisphere.  We tested the
significance of this feature using the following procedure:

\begin{enumerate}
\item Fit the `true' map using the observed data (Fig.~\ref{fig:amherdisp}).
\item Compute 200 `trial' maps from simulated data sets constructed
using bootstrap re-sampling. From our observed trailed spectrum
containing $n$ data points we formed a simulated trailed spectrum by
selecting, at random and with replacement, $n$ data values and placing
these at their original positions in the new simulated trailed
spectrum. For points that are not selected, the associated error bar
was set to infinity and hence they were effectively omitted from the
fit. For points selected more than once, the error bars were divided
by the square root of the number of times they were picked (see
Paper I for more details).
\item Calculate 200 `difference' maps by subtracting each `trial' map from
the `true' map. These `difference' maps give the scatter in each pixel.
(It would be incorrect to now calculate a summary error statistic like
the standard deviation, as the distribution of pixel values is often
found to be non-normal; see Paper I).
\item Subtract some mean or comparison level from the `true' map. For
instance, if one wished to determine the significance of a starspot, an
appropriate comparison level to subtract from the `true' map would be a
region of immaculate photosphere. Subtract the corresponding level from
each of the `difference' or `scatter' maps.
\item Rank each pixel in the `true' map computed in step (iv) against the
corresponding pixel in the `scatter' maps. This will show the fraction
of fluctuation values that the observed value in the `true' map exceeds.
\end{enumerate}

\begin{figure*}
\psfig{figure=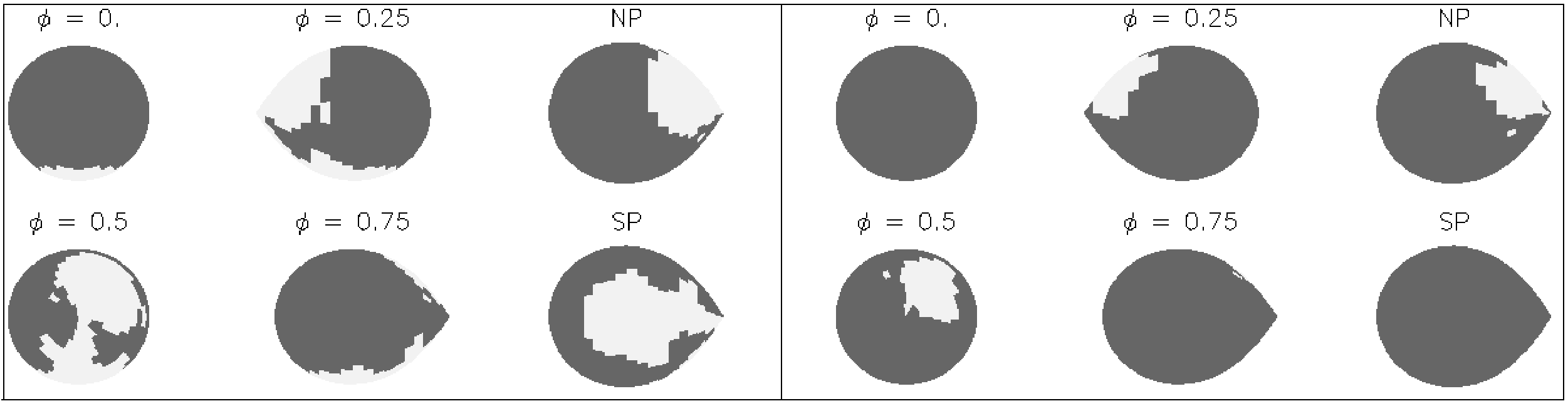,width=18.cm,angle=0.}
\caption{Significance maps of AM Her. The four images on the left-hand
side of each panel are viewed in the equatorial plane, and the two
images on the right-hand side of each panel are viewed from above and
below the north and south pole respectively. The significance map in
the left-hand panel was constructed by taking the comparison level to be the
average intensity of the outer hemisphere. The significance map in the
right-hand panel was constructed by taking the comparison level to be the
average intensity of the region around the $L_1$ point that does not
appear to be irradiated.  Bright greyscales indicate regions where the
`true' map ranks above 100 per cent of the `scatter' maps. The
significance of the region near the south pole (SP) in the left-hand
panel is an artefact, since this region is never observed.}
\label{fig:amhererr1}
\end{figure*}

For our first test we took the average intensity of the outer (and
presumably non-irradiated) hemisphere of the star as our comparison
level, thereby testing the reality of the asymmetry between the outer
and inner hemispheres.  The left-hand panel in Fig.~\ref{fig:amhererr1}
shows the resultant significance map, calculated using the procedure
listed above. The bright greyscale depicts regions where the `true'
map ranks above 100 per cent of the `scatter' maps, and therefore
represents regions where the intensity is significantly different from
the outer hemisphere. Dark greyscales represent regions of the map
that do not rank highly against the scatter map. Features in these
regions are not, therefore, significantly different from the average
intensity of the outer hemisphere.

The left-hand panel in Fig.~\ref{fig:amhererr1} shows that the irradiation
seen on the inner hemisphere in the Roche tomogram is real, and
appears more significant on the trailing hemisphere (i.e., $\phi$ =
0.25).  Note that the significance of the regions near the south pole
(SP) is an artefact.  Due to the inclination of AM Her, latitudes $>
i$ in the southern hemisphere are never seen and hence the
reconstructions always assume the (same) default level, with very
little scatter in the intensity values. As a result, regions near the
south pole will appear significantly different from the comparison
level unless, of course, the comparison level is very similar to the
default level.

The right-hand panel in Fig.~\ref{fig:amhererr1} shows another
significance map, but taking the comparison level as a region around
the $L_1$ point on the leading hemisphere ($\phi$ = 0.25). This shows
that the asymmetry between the leading and trailing hemispheres is
real, and that the effects of irradiation are primarily found on the
trailing hemisphere. The asymmetry in the irradiation pattern between
the northern and southern hemisphere is again an artefact, since
regions on the southern hemisphere are less visible and the data does
not constrain the reconstructions in these regions. Note that in this
case, the comparison level selected was similar to the default level
of the map, and therefore regions near the south pole (SP) no longer
appear to be significant.

Our results are in agreement with the surface maps of AM Her obtained
using the method of radial-velocity curve fitting
(\citealt{davey92,davey96}). They proposed that the lack of
irradiation on the leading hemisphere could be explained if the
incoming accretion column was able to block the radiation produced
close to the white dwarf and found that the accretion geometry given
by \citet{cropper88a} supported this argument.

\section{Results for QQ Vul}
\label{sec:qqvul}

\subsection{The binary geometry of QQ Vul}
The systemic velocity and inclination of QQ Vul were determined in the
manner outlined in Section~\ref{sec:parameters}, and the results are
shown in Fig.~\ref{fig:incl}. We find that the systemic velocity is
well constrained with a value of --14 km s$^{-1}$, comparable with
the value of 17 $\pm$ 23 km s$^{-1}$ obtained by
\citet{mukai87}. \citet{catalan99} found $\gamma$~=~--0.2 $\pm$ 2.2 km
s$^{-1}$ from the results of elliptical fits to the Na I absorption
lines used in this work. In addition, using the same method they also
obtained $\gamma$ = --8.3 $\pm$ 1.6 km s$^{-1}$ using data obtained on
the WHT in 1993.  One explanation for this difference in the systemic
velocity is if a change in the intensity distribution between the two
epochs can alter the measured value of $\gamma$ (we explore this
possibility in Section~\ref{sec:huaqr}).

It is clear from Fig.~\ref{fig:incl} that it is difficult to select a
single value for the inclination given the scatter in the distribution
of the points. The quality of the reconstructions does, however, begin
to deteriorate towards lower inclinations allowing us to set a lower
of limit of $\sim$72$^{\circ}$ for the inclination of QQ Vul.

Previous estimates of the orbital inclination include $i$~=
50$^{\circ}$--70$^{\circ}$ \citep{schwope00}, $i$ =
65$^{\circ}\pm$7$^{\circ}$ \citep{catalan99}, $i$ = 60$^{\circ} \pm$
14$^{\circ}$ \citep{mukai87} and 46$^{\circ}<i<$ 74$^{\circ}$
\citep{nousek84}. In addition, \citet{nousek84} also found evidence of
a brief eclipse of the cyclotron region combined with the simultaneous
disappearance of emission lines. They attributed this to an eclipse by
the secondary star, forcing the orbital inclination to the upper limit
of 46$^{\circ}<i<$ 74$^{\circ}$. From EUVE observations
\citet*{belle00} also suggested that the X-ray emitting region is
eclipsed by the secondary star, constraining the inclination to
$>$~60$^{\circ}$. Certainly, in this work we can rule out inclinations
$<$~60$^{\circ}$ due to the low entropy of the reconstructions at
these inclinations. These claims of eclipse features, however, are
outside phase 0 and thus cannot be attributed to the secondary star.
The lack of an X-ray eclipse at this phase means that an inclination
$>$ 72$^{\circ}$ can be excluded.

Fig.~\ref{fig:landscapes} shows the entropy landscape for QQ Vul,
constructed assuming an inclination of 72$^{\circ}$, from which we
derive optimum masses of $M_1$ = 0.66 M$_{\odot}$ and $M_2$ = 0.42
M$_{\odot}$. We adopt these masses and inclination for the
reconstruction in Section~\ref{qqvulmap}. The derived masses, however,
decrease smoothly as the inclination is increased to 90$^{\circ}$ and
cover the range $M_1$ = 0.58--0.66 M$_{\odot}$, $M_2$ = 0.34--0.44
M$_{\odot}$ at a constant $q$ = 0.63. This agrees well with,
and is better constrained than, the masses of $M_1$ =
0.54$^{+0.21}_{-0.16}$ M$_{\odot}$ and $M_2$~=~0.30$\pm$0.10
M$_{\odot}$ estimated by \citet{catalan99}.

The white-dwarf mass we derive agrees with the values of $M_1$ =
0.58$^{+0.44}_{-0.09}$ M$_{\odot}$ \citep{mukai87} and
$M_1$~=~0.59~M$_{\odot}$ \citep{mouchet93}, but is almost half that of
\citet{cropper98} and \citet{wu95} who obtained $M_1$ = 1.22
M$_{\odot}$ and $M_1$~=~1.1--1.3 M$_{\odot}$, respectively. The two
latter mass determinations make use of the X-rays emitted from the
accreting white dwarf and involve fitting model spectra to the
observed X-ray spectra, and are therefore subject to underlying
assumptions in the model.

The mass ratio of $q$ = 0.63 obtained in this work is in
excellent agreement with previous determinations, including $q$ =
0.54$\pm$0.14 \citep{catalan99} and $q$ = 0.45--0.67 determined by
\citet{schwope00} from Doppler tomography and polarimetry.

\subsection{The surface map of QQ Vul}
\label{qqvulmap}

\begin{figure*}
\psfig{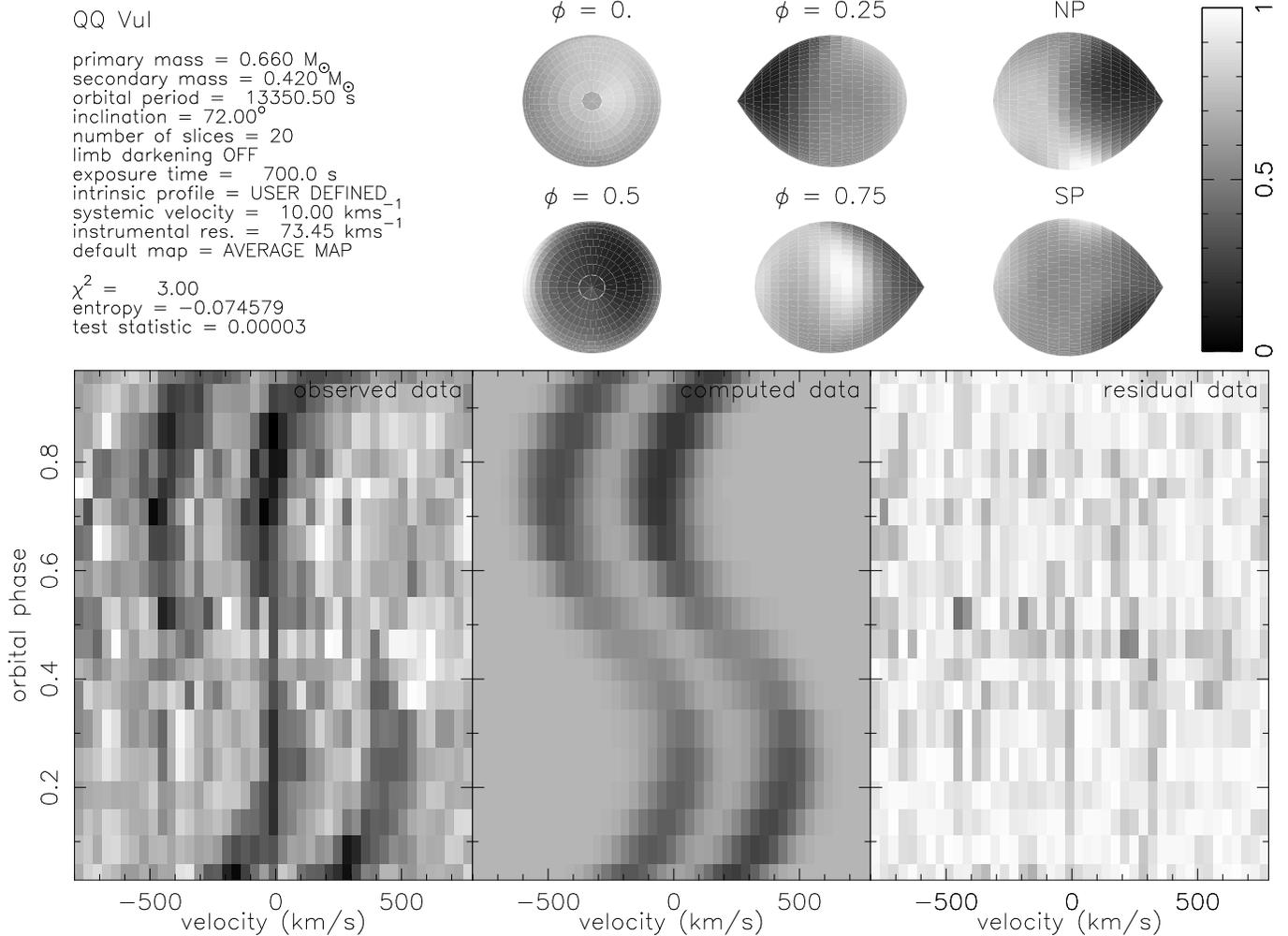}
\caption{Roche tomogram of QQ Vul. Bright greyscales in the tomograms
depict regions of enhanced Na I absorption. The greyscales are
reversed in the trailed spectra.  Note that the systemic velocity
quoted in the upper-left of the panel does not take into account the
systemic velocity of the spectral-type template star.}
\label{fig:qqvulrdisp}
\end{figure*}

The Roche tomogram of QQ Vul is presented in Fig.~\ref{fig:qqvulrdisp}
and shows a distinct reduction in the Na I absorption around the $L_1$
point which can be explained by irradiation from the white dwarf and
accretion regions ionising the Na I on the inner face of the
secondary, as in AM Her.  The effects of irradiation also appear to be
slightly stronger on the trailing hemisphere, suggesting that the
leading hemisphere may be partially shielded by the accretion stream
or curtain.

Another notable feature is a bright patch on the leading hemisphere,
corresponding to an increased Na I flux deficit around phase 0.75.  A
similar feature can also be seen in the tomogram of IP Peg
(Fig.~\ref{fig:ippegrdisp}) and larger flux deficits at $\phi$ = 0.75
than at $\phi$ = 0.25 have also been reported for the Na I doublet in
HT Cas \citep*{catalan99b} and in the TiO light curves of Z Cha
\citep{wade88}.

In their study of QQ Vul using the same data used in this work,
\citet{catalan99} found that their surface maps derived from
radial-velocity curve fitting did not give satisfactory fits to the
velocities and line fluxes between orbital phases $\phi$ = 0.6 and
$\phi$ = 0.8. The bright patch seen around phase 0.75 in
Fig.~\ref{fig:qqvulrdisp} is almost certainly the cause of their
unsatisfactory fits. \citet{catalan99} attributed this discrepancy to
the presence of starspots. This interpretation is almost certainly
wrong, however, as an increase in the Na I flux deficit is
inconsistent with a star-spot, as the Na I flux deficit is known to
decrease with later spectral type \citep{brett93} and star-spots are
generally cooler than the surrounding photosphere. Therefore, a
starspot should appear as a dark feature in the tomograms. (Although
spots hotter than the photosphere have also been imaged, e.g.
\citealt{donati92}).

\begin{figure}
\psfig{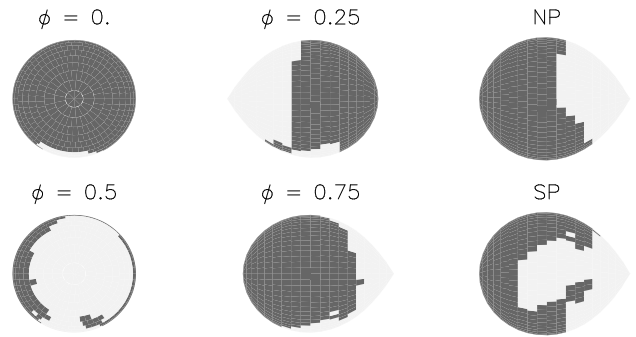}
\caption{The significance map of QQ Vul, taking the comparison level
to be the average intensity of the outer-hemisphere, avoiding
both the irradiated regions and the bright area on the leading
hemisphere. Bright greyscales indicate regions where the `true' map
ranks above 100 per cent of the `scatter' maps. The significance of
the region near the south pole (SP) is an artefact since this region
is never observed.}
\label{fig:qqvulerr1}
\end{figure}

The reality of the spot feature can be assessed in
Fig.~\ref{fig:qqvulerr1}, which shows the significance map when we
take the comparison level as the average intensity on the outer
hemisphere, avoiding both the region that appears to be irradiated and
the bright patch.  This shows that the bright patch is not
significant. The significance map does, however, show that the
irradiated inner hemisphere is real and there is evidence for stronger
irradiation of the trailing hemisphere.

As with AM Her, the reduction in absorption around the $L_1$ point can
be explained by irradiation from the white dwarf and accreting regions
ionising the Na I on the inner face of the secondary, with possible
shielding by the accretion stream/curtain reducing the effect on the
leading hemisphere.

\section{Results for IP Peg}

\subsection{The binary geometry of IP Peg}
\label{sec:ippeg}

The optimum systemic velocity, as shown in Fig.~\ref{fig:incl}, is
--9~km~s$^{-1}$. This is significantly different from the
values of 56 $\pm$ 4 km s$^{-1}$ obtained by \citet*{martin87} and
31 $\pm$ 7~km~s$^{-1}$ obtained by \citet{martin89}.
Fig.~\ref{fig:incl} also shows that we have been unable to determine
the inclination of IP Peg. This is due to the small variations in the
line strength over the orbital cycle, which provides our only
constraint on the inclination before taking into consideration the
eclipse width. The reconstructions do, however, yield a consistent
result of 0.43 for the mass ratio of IP Peg
(Fig.~\ref{fig:mratio}).  This is in disagreement with
$q$~=~0.55--0.63 and $q$~=~0.55--0.62 found by \citet{martin89} and
\citet{marsh88a},  but agrees with 0.35$<$q$<$0.49 \citep{wood86},
$q$~=~0.39$\pm$0.04 \citep{catalan99b} and $q$~=~0.32$\pm$0.08
\citep{beekman00}. All the authors, except \citet{beekman00}, made
corrections for the effects of irradiation where appropriate.

Although we cannot determine the inclination using the surface maps
alone, IP Peg is an eclipsing system, and its geometry is further
constrained by the observed eclipse width, which is dependent upon the
inclination and size of the secondary star's Roche
lobe. Fig.~\ref{fig:landscapes} shows the entropy landscape
constructed by varying the inclination at each different mass pairing
in order to match the eclipse width of $\Delta\phi$ = 0.0863
\citep{wood86}. This results in optimum masses of $M_1$ = 1.18
M$_{\odot}$ and $M_2$~=~0.50~M$_{\odot}$ at an inclination of
82.3$^{\circ}$. Altering the eclipse width to the other extreme of
$\Delta\phi$ = 0.0918 given by \citet{wood86} changes this
determination to $M_1$ = 1.16 M$_{\odot}$ and $M_2$ =
0.50 M$_{\odot}$ and the optimum inclination to
84.4$^{\circ}$. Our best estimates are therefore $M_1$ = 1.16 -- 1.18
M$_{\odot}$, $M_2$~=~0.50~M$_{\odot}$ and $i$ =
82$^{\circ}$--85$^{\circ}$.

Our secondary star mass determination appears to lie between the
values of $M_2$ = 0.59--0.75 M$_{\odot}$ \citep{martin89},
$M_2$~=~0.55--0.73~M$_{\odot}$ \citep{marsh88a} and
$M_2$~=~0.33$^{+0.14}_{-0.05}$~M$_{\odot}$ \citep{beekman00},
$M_2$~=~0.33$\pm$0.07~M$_{\odot}$ \citep{catalan99b}.

The primary mass is consistent with the values of $M_1$~=~1.05--1.25
M$_{\odot}$ determined by \citet{martin89}, $M_1$ = 0.99--1.19
M$_{\odot}$ \citep{marsh88a} and
$M_1$~=~1.05$^{+0.14}_{-0.07}$~M$_{\odot}$ \citep{beekman00}, but
slightly larger than $M_1$~=~0.85$\pm$0.09~M$_{\odot}$ determined by
\citet{catalan99b}.

\subsection{The surface map of IP Peg}
\label{ippegmap}

\begin{figure*}
\psfig{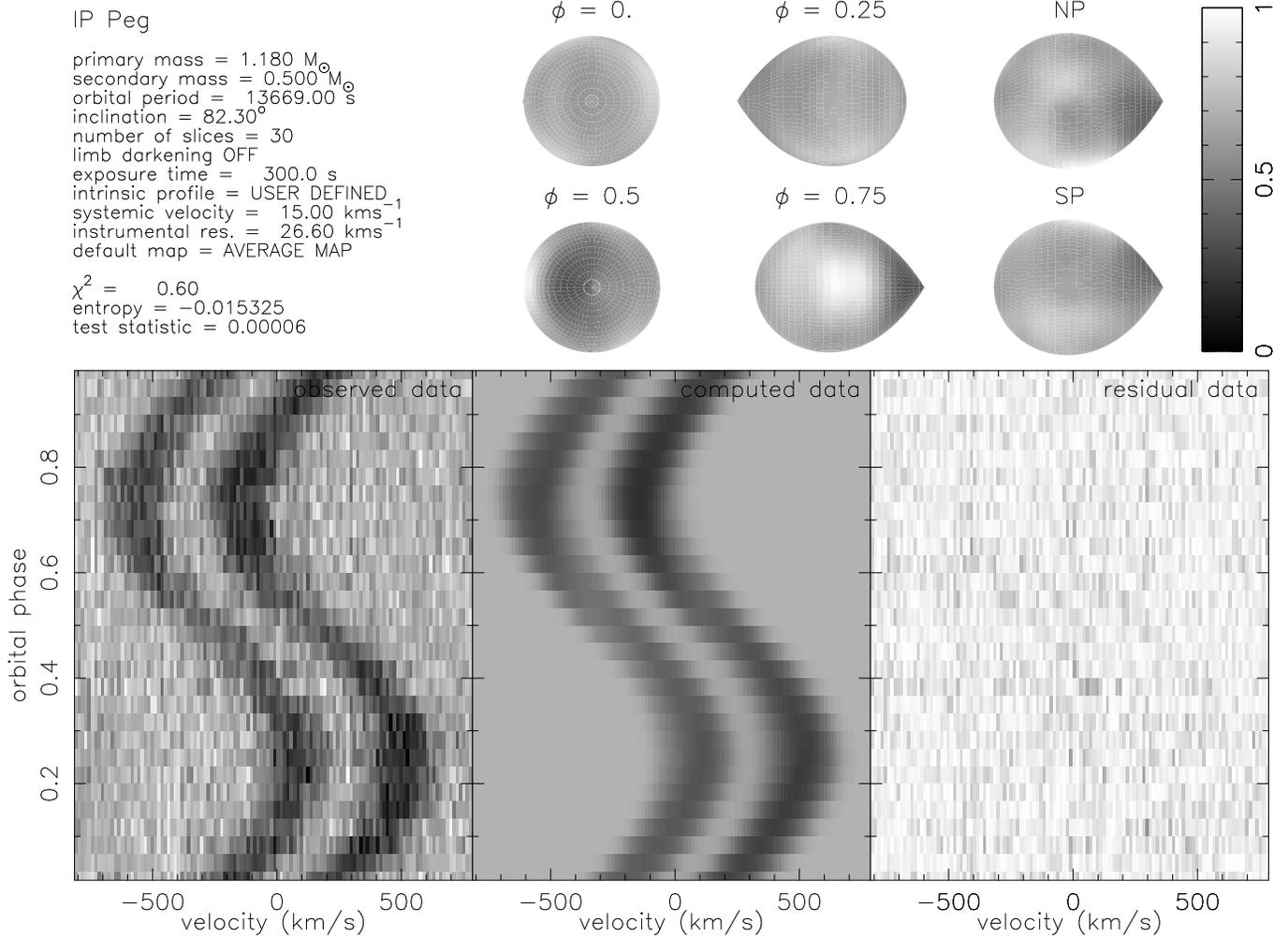}
\caption{Roche tomogram of IP Peg. Bright greyscales in the tomograms
depict regions of enhanced Na I absorption. The greyscales are
reversed in the trailed spectra.  Note that the systemic velocity
quoted in the upper-left of the panel does not take into account the
systemic velocity of the spectral-type template star.}
\label{fig:ippegrdisp}
\end{figure*}

The Roche tomogram of IP Peg in the light of the Na I absorption
doublet is shown in Fig.~\ref{fig:ippegrdisp}. The most noticeable
feature is the weakened absorption flux deficit on the inner
hemisphere of IP Peg, which can be attributed to irradiation.  The
effects of irradiation appear to be strongest, however, towards the
leading hemisphere of the secondary star. We also see a bright patch
of enhanced Na I absorption on the leading hemisphere similar to that
seen in QQ Vul (Section~\ref{qqvulmap}).

\begin{figure}
\psfig{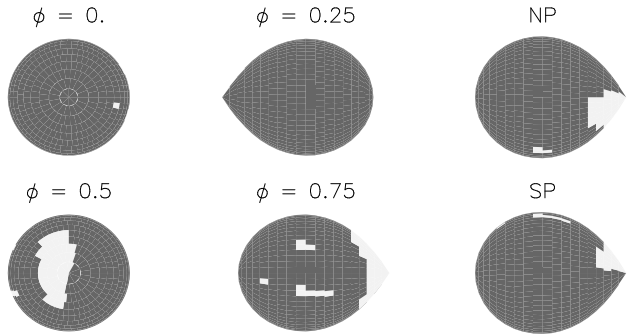}
\caption{The significance map of IP Peg, taking the comparison level
to be the average intensity of the outer-hemisphere avoiding
the bright region on the leading hemisphere. Bright greyscales
indicate regions where the `true' map ranks above 100 per cent of the
`scatter' maps.}
\label{fig:ippegerr1}
\end{figure}

The significance map constructed taking the comparison level to be the
average intensity of the outer hemisphere, avoiding the bright patch,
is shown in Fig.~\ref{fig:ippegerr1}. From this we can see that the
bright patch is not significant, whereas the irradiated region on the
leading hemisphere is.

In their radial velocity curve-fitting studies, \citet{davey92} also
found that the effects of irradiation were strongest on the leading
hemisphere, and that regions of IP Peg appeared to be irradiated some
distance onto the back half of the star. They found that this feature
could not be explained by irradiation from the bright spot, as the
illuminating source would have to originate from a direction about
45$^{\circ}$ away from the line of centres and instead attributed this
asymmetry to circulation currents (e.g. \citealt{martin95}).

It is clear in Fig.~\ref{fig:ippegrdisp} that we cannot identify any
significant irradiation effects on the back half of the star and, as
such, cannot provide any support for the evidence of circulation
currents on IP Peg.  It seems likely, instead, that the irradiation of
the leading hemisphere can be explained by the bright spot, which is
located on the correct side of the secondary star to cause this
effect, and is a strong illuminating source (as seen in, for example,
\citealt{szkody87}).

Our Roche tomogram, however, bears little resemblance to the surface
map that \citet{davey92} computed from a model for a secondary star
heated equally by the bright spot and the white dwarf (their Fig. 15),
in that there does not appear to be any significant signs of
irradiation on the trailing hemisphere. This suggests that irradiation
from both the white dwarf and the bright spot is shielded from this
region. An obvious way to block irradiation from the surface of the
white dwarf is through shielding by the accretion disc.  Irradiation
from the bright spot may be blocked by the accretion stream,
especially if much of the emission from the bright spot is located
downstream from the impact point, as suggested in the
eclipse maps of IP Peg by \citet{bobinger99}. This would result in
shielding of the trailing hemisphere, but still allow the bright spot
to irradiate the leading hemisphere.

One further explanation for the features seen around the $L_1$ point
is that they could be caused by the effects of an eclipse of the
secondary star by the accretion disc. This would cause darker regions
to appear in the Roche tomogram around the $L_1$ point and a reduction
in the quality of the fit.  Reconstructions carried out neglecting
phases between 0.45--0.55 show no improvement in the quality of the
fit, and there is negligible difference in the Roche tomograms.  We
are therefore confident that these features are not the result of an
eclipse.

\section{Results for HU Aqr}
\label{sec:huaqr}

\subsection{The binary geometry of HU Aqr}
\label{sec:huaqrgeo}
Fig.~\ref{fig:incl} shows the systemic velocity and inclination of HU
Aqr as a function of entropy. We obtain a value of $\gamma$ = 7
km s$^{-1}$ and, although it is difficult to assign a value to the
inclination, the reconstruction quality begins to decrease rapidly for
inclinations less than 80$^{\circ}$, and we therefore adopt a best
estimate of $i\geq$ 80$^{\circ}$.

Our estimate of the systemic velocity is compatible at the 2$\sigma$
level with $\gamma$ = 83 $\pm$ 44 km s$^{-1}$ derived by \citet{glenn94}
who fitted a sinusoid to the narrow H$\alpha$ component thought to
originate from the secondary star. \citet{schwope97} derived a
systemic velocity of --7.8 $\pm$ 1.4 km s$^{-1}$ by fitting radial
velocity curves to the same data used in this work. Although this
result disagrees with the value obtained using Roche tomography, the
sine-fitting method takes no account of any non-uniform intensity
distribution present on the secondary star, whereas Roche tomography
does. Since there is clearly an asymmetrical intensity distribution on
the secondary star in HU Aqr (see Section~\ref{huaqrmap}) we believe
that this is the cause of the difference between the two results.

In order to test the reliability of the sine-fitting method in
determining the systemic velocity, we generated a model secondary star
with the same binary geometry as we have found for HU Aqr in this work
($M_1$ = 0.61 M$_{\odot}$, $M_2$~=~0.15~M$_{\odot}$ and $i$ =
84.4$^{\circ}$, see later) but with $\gamma$ = 0 km s$^{-1}$. The
inner hemisphere of the model was then uniformly irradiated except for
a totally dark shadow covering 40 per cent of the leading
hemisphere. The outer hemisphere of the secondary star was also set to
be totally dark. From this model, trailed spectra of infinite
signal-to-noise were generated, assuming 100 equally spaced phase bins
and 30 second exposure lengths.

\begin{figure}
\psfig{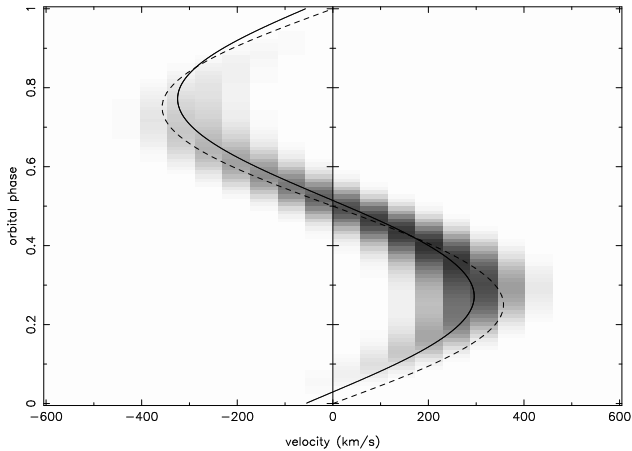}
\caption{The synthetic trailed spectrum used to test the reliability
of the sine-fitting method for the purposes of determining the
systemic velocity.  See the text for a description of the model used
to generate the synthetic data (Section~\ref{sec:huaqrgeo}).  The
solid line corresponds to the radial-velocity curve fit and the dashed
line corresponds to the true motion of the centre-of-mass of the
secondary star. Note that the radial-velocity curve fit has also
adjusted the zero point to $\phi$ = 0.02.}
\label{fig:rvcurve}
\end{figure}

By fitting radial-velocity curves to this fake data, we obtained a
systemic velocity of --14 km s$^{-1}$. Fig.~\ref{fig:rvcurve} shows
the result of the radial-velocity curve fit (solid line) to the
synthetic trailed spectrum, and the true motion of the centre-of-mass
of the secondary star (dashed line) for comparison. The value of
$\gamma$ = --14 km s$^{-1}$ is significantly different from the true
value of $\gamma$ = 0 km s$^{-1}$ used in the model, and brings our
systemic velocity of 7 km s$^{-1}$ into agreement with the value of
--7.8 $\pm$ 1.4 km s$^{-1}$ obtained by \citet{schwope97}.  This shows
that a knowledge  of the intensity distribution across the secondary
is required to accurately determine the systemic velocity (as also
found by \citealt{schwope97}). This may have implications for the
accurate determinations of $\gamma$'s used in CV age tests (e.g.,
\citealt*{vanparadijs96} and \citealt{north02}).

Since HU Aqr is an eclipsing system, its inclination has been
reasonably constrained by previous studies, including estimates of $i$
= 80$^{\circ}\pm$5$^{\circ}$ \citep{glenn94} obtained from fits to
polarisation curves, and $i$ = 85.6$^{\circ}$ \citep{schwope01} from
modelling the soft X-ray light-curves. Our determination of $i\geq$
80$^{\circ}$ does not provide any further constraint to those values
already published.

Previous estimates of the mass ratio, however, are somewhat more
uncertain and range from roughly 0.18 -- 0.4 (e.g. \citealt{schwope97},
\citealt{hakala93} and \citealt*{heerlein99}).  Fig.~\ref{fig:mratio} shows
the component masses and mass ratio obtained when reconstructions are
carried out over different inclinations. This yields a mass ratio for
HU Aqr of $q$ = 0.25, which is in excellent agreement with
$q$ = 0.256 derived by \citet{heerlein99} assuming $i$ = 84$^{\circ}$,
and the value of $q$~=~0.25 that \citet{schwope01} found.

Fig.~\ref{fig:landscapes} shows the entropy landscape for HU Aqr which
has been constructed by changing the inclination at each different
mass pairing in order to match the eclipse width of $\Delta\phi$ =
0.0763 \citep{heerlein99}, as described for IP Peg in
Section~\ref{sec:ippeg}. This results in a best estimate of the
component masses of $M_1$ = 0.61 M$_{\odot}$ and $M_2$ =
0.15 M$_{\odot}$. This agrees with the result of \citet{schwope01},
who determined $M_2$ = 0.15 M$_{\odot}$ using the mass-radius relation of
\citet{caillaut90}, and $M_2$ = 0.17 M$_{\odot}$ according to the
relation by \citet{neece84}.
Our derived masses, combined with
the eclipse width, give our best estimate of the inclination as
84.4$^{\circ}$.

\subsection{The surface map of HU Aqr}
\label{huaqrmap}

\begin{figure*}
\psfig{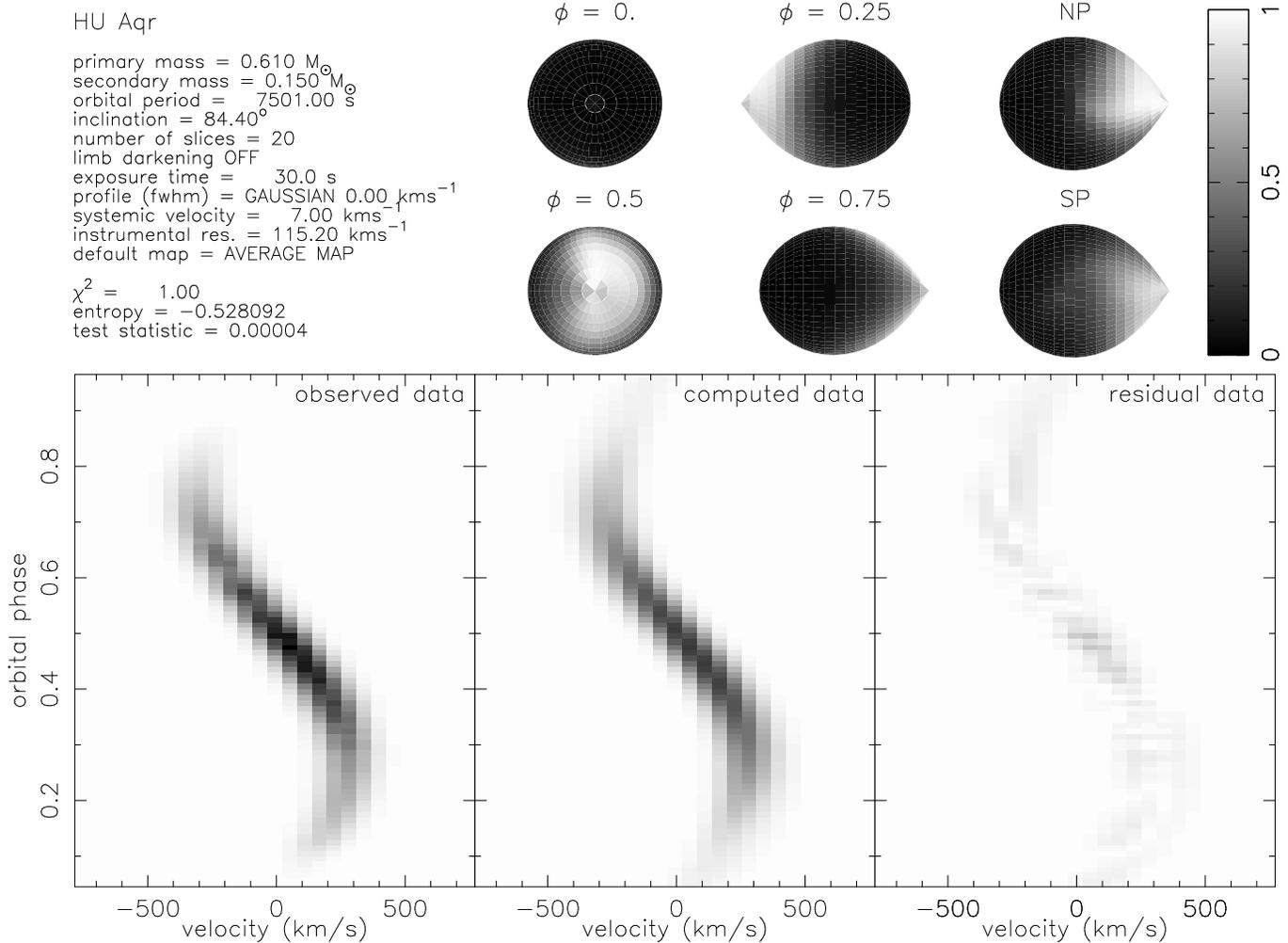}
\caption{Roche tomogram of HU Aqr. Bright greyscales in the tomograms
depict regions of enhanced He II $\lambda$4686\AA~emission. The greyscales
are reversed in the trailed spectra.}
\label{fig:huaqrrdisp}
\end{figure*}

The Roche tomogram of HU Aqr in the light of the He II
4686\AA~emission line is presented in Fig.~\ref{fig:huaqrrdisp}. The
two most prominent features in the Roche tomogram are the strong
asymmetry between the inner and outer hemispheres of the secondary
star and the weaker asymmetry between the leading and trailing
hemispheres. The emission from the inner hemisphere is due to
irradiation from the primary and the accretion regions, and the
tomogram is compatible with 90 per cent of the emission arising from
the inner hemisphere alone. (Note that the remainder of the emission
that  arises on the outer hemisphere is most likely due to the
smearing effects of the default map).

\begin{figure*}
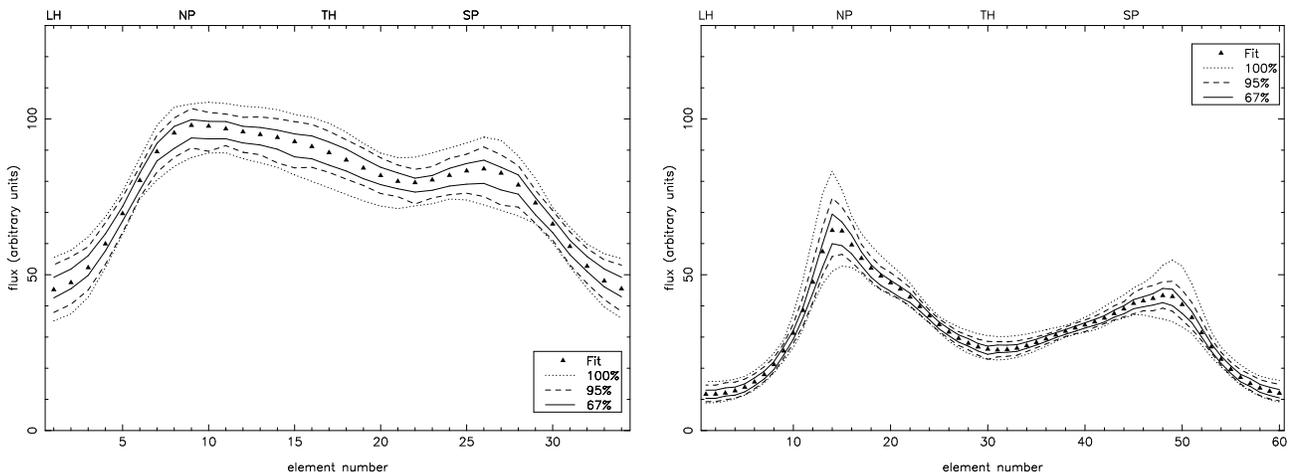

\begin{tabular}{ll}
\psfig{figure=fig12a.ps,width=8.3cm,angle=-90} &
\psfig{figure=fig12b.ps,width=8.3cm,angle=-90} \\
\end{tabular}
\caption{The left-hand panel shows a vertical slice (see text for
explanation) through HU Aqr near the $L_1$ point; the right-hand panel
shows a vertical slice through HU Aqr closer to the terminator. The
triangular points show the intensities found in the original fit. The
lines show confidence limits derived from 200 bootstrapped
reconstructions. The labels on the top axis denote the positions of
pixels located on the leading hemisphere (LH), north pole (NP),
trailing hemisphere (TH) and south pole (SP).  }
\label{fig:huaqrslices}
\end{figure*}

The reality of the asymmetry between the leading and trailing
hemispheres can be assessed from the left-most panel in
Fig.~\ref{fig:huaqrslices}, which shows a slice passing through the
leading hemisphere (LH), north pole (NP), trailing hemisphere (TH) and
south pole (SP) near the $L_1$ point of the secondary star.  The
triangular points represent the `true' map values; 67 per cent of the
bootstrapped `trial' maps (measured relative to the mode of the
distribution) lie within the solid line, and 100 per cent within the
dotted line\footnote{We have not generated a significance map for HU
Aqr since the low scatter in the pixel values derived from the
bootstrap reconstructions mean that each region of the map would be
significant when compared with any other region.}. It is clear from
this that the asymmetry between the leading and trailing hemispheres
is not due to statistical errors. This asymmetry was also found by
\citet{schwope97} who attributed it to shielding by the accretion
curtain.

Closer examination of the tomogram reveals more subtle intensity
variations across the secondary star. In particular there appears to
be a region of more intense emission located towards the north pole
but shifted slightly towards the leading hemisphere. The plot on the
right hand-side of Fig.~\ref{fig:huaqrslices} shows a vertical slice
cutting through the secondary star close to the polar regions but
slightly towards the $L_1$ point. This again shows the shadow on the
leading hemisphere (LH) as discussed earlier, but another feature of
note is the drop in emission from the trailing hemisphere (TH)
compared to the polar regions. This feature is not apparent near the
$L_1$ point and only becomes significant for regions halfway between
the centre-of-mass of the secondary and the $L_1$ point, disappearing
again when the effects of irradiation are diminished near the
terminator. This cannot be explained by shielding of the secondary
star by the accretion curtain, since this would only block irradiation
on the leading hemisphere.

HU Aqr does, however, have a strong stream component with as much as
40 per cent of the flux in the U band due to accretion stream emission
\citep{hakala93}. Doppler maps by \citet{schwope97} also show that
the accretion stream is a dominant emission-line feature.  Presumably
the accretion stream can also irradiate the secondary star and, since
the stream is deflected in the direction of the orbital motion, the
stream will be invisible to regions of the secondary star on the
trailing hemisphere away from the $L_1$ point. This obscuration of the
accretion stream by the secondary star itself may explain the observed
shadow on the trailing hemisphere.

This feature could also be explained if systematic errors were able to
produce a greater line flux in the centres of the lines, corresponding
to artificially brighter polar regions in the Roche tomograms and
giving the illusion of a shadow at lower latitudes. It was first
thought that the assumption of a Gaussian profile for the emission
from the secondary star could introduce such an artefact, since the
actual profile from the secondary star is expected to be flatter,
especially around quadrature. We found, however, that at low spectral
resolutions the emission profile from the secondary is well
approximated by a Gaussian.  Artefacts could be generated if the
emission from the various components of HU Aqr (e.g. from the
secondary star and the accretion stream) are incorrectly disentangled,
which is especially difficult at phases where the individual
components are crossing each other. This situation is more difficult
to assess and emphasises the importance of mapping lines that are
known to originate solely from the secondary star.

\section{Conclusions}
The Roche tomograms show an asymmetric irradiation pattern on all 4
CVs.  This can be explained if the accretion column/stream shields
the leading hemisphere in AM Her, QQ Vul and HU Aqr. There is also
evidence for a fainter shadow present on the trailing hemisphere of HU
Aqr. We propose that the stream itself contributes significantly to
the irradiation, and that the shadow on the trailing hemisphere occurs
as these regions on the secondary star have no direct view of the
majority of the stream. Further observations using absorption lines
originating from the secondary star are needed to confirm this.

A different scenario occurs in IP Peg, where irradiation affects the
leading hemisphere only. We suggest that the bright spot, which is on
the correct side of the secondary star to cause such an irradiation
pattern, may be the principal illuminating source in this
system. Although we still expect irradiation of the trailing
hemisphere by the white dwarf (as seen in the magnetic CVs), there is
no evidence for this in the tomogram. It may be the case that the
accretion disc in IP Peg is shielding the secondary star from
irradiation by the white dwarf. Unlike \citet{davey92}, we find no
evidence for irradiation extending onto the back half of the star, and
hence cannot support their idea that circulation currents operate in
IP Peg.

We have used the entropy landscape technique to derive the system
parameters.  Although we cannot reliably constrain the inclination,
and therefore the masses, in the non-eclipsing CVs, the technique does
provide a consistent result for the mass ratio that is independent of
the inclination. For the two eclipsing CVs (IP Peg and HU Aqr), we
believe that the masses derived in this work are the most accurate to
date. In addition, it is found that measurements of the systemic
velocity using circular orbit fits to the radial-velocity variations
are unreliable, and are dependent upon the intensity distribution
across the secondary star. We might therefore expect such measurements
to be both line and time-dependent. Since the intensity distribution
across the secondary star is known in Roche tomography, systemic
velocity measurements using this technique should prove far more
accurate.

\section*{\sc Acknowledgements}
We thank Stuart Littlefair for many useful conversations. CAW is
supported by a PPARC studentship. This project was supported in part
by the Bundesministerium f\"ur Bildung und Forschung through the
Deutsches Zentrum f\"ur Luft- und Raumfahrt e.V. (DLR) under grant
number 50 OR 9706 8. We would also like to thank the referee, Andrew
Cameron, for his comments, which substantially improved the clarity of
the paper.

\newpage
\bibliographystyle{mn2e}
\bibliography{abbrev,refs}

\end{document}